\documentclass[11pt,a4paper]{article}
\makeatletter
\let\@fnsymbol\@arabic
\makeatother
\usepackage[T1]{fontenc}
\usepackage[utf8]{inputenc}
\usepackage{authblk}
\usepackage{longtable}
\usepackage[hyphens]{url}
\usepackage{graphicx}
\usepackage{pdflscape}
\usepackage{float}
\usepackage{color}
\usepackage[stable]{footmisc}
\usepackage[english]{babel} 
\usepackage{ifthen}
\usepackage{fancyhdr}
\title{Online Social Networks: Threats and Solutions}
\author[*]{Michael Fire\thanks{mickyfi@bgu.ac.il}}
\author[**]{Roy Goldschmidt\thanks{rgoldshmidt@knesset.gov.il}}
\author[*]{Yuval Elovici \thanks{elovici@bgu.ac.il}}
\affil[*]{Telekom Innovation Laboratories  at Ben-Gurion University of the Negev}
\affil[**]{The Knesset Research and Information Center}

\newcommand{\s}[1]{{#1}}
\date{}

\begin{document}
\maketitle

\begin{abstract}

Many online social network (OSN) users are unaware of the numerous security risks that exist in these networks, including privacy violations, identity theft, and sexual harassment, just to \s{name} a few. According to recent studies, \s{OSN} users readily expose personal and private details about themselves, such as relationship status, date of birth, school name, email address, phone number, and even home address. This information, if \s{put} into the wrong hands, can be used to harm users both in the virtual \s{world} and in the real world. These risks become even more severe when the users are children. 
In this paper we present a thorough review of the different security and privacy risks which threaten the well-being of OSN users in general, and children in particular. In addition, we present an overview of existing solutions that can provide better protection, security, and privacy for OSN users. 
\s{We also} offer simple-to-implement recommendations for OSN users which can \s{improve} their security and privacy when using these platforms. \s{Furthermore, we suggest future research directions. }
\\\\

\noindent \textbf{Keywords.} Online Social Networks, Security and Privacy, Online Social Network Security Threats, Online Social Network Security Solutions.
\\\\
\textbf{To appear in:} ``\textit{IEEE Communications Surveys \& Tutorials}.''

\end{abstract}

\section{Introduction}
\begin{figure}[h]

\begin{center}
\includegraphics[scale=0.5,angle=-90]{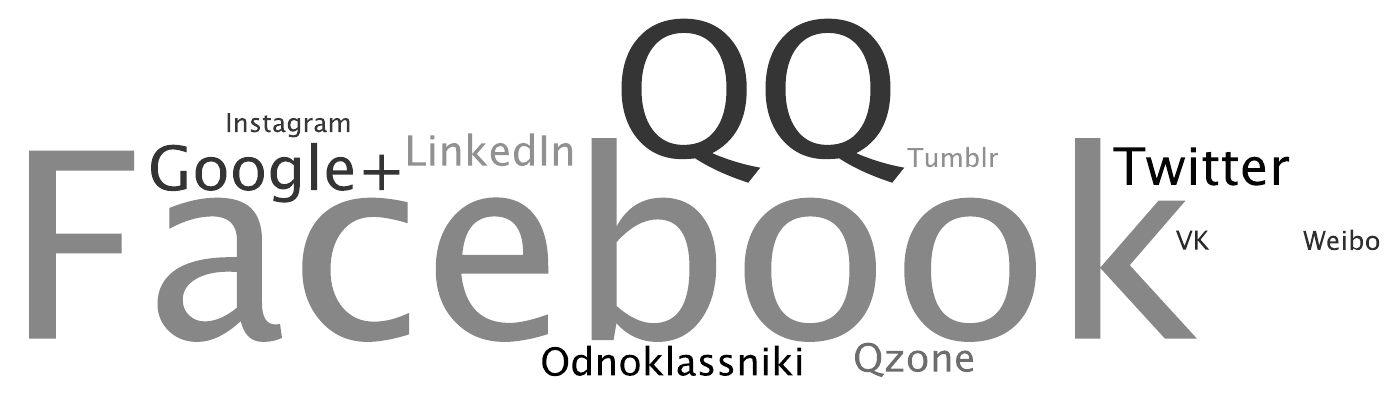}
\end{center}
\caption{\textbf{Word Cloud of OSNs with More Than  100 Million Active Users.} This word cloud was constructed using Wordle~\cite{wordle} where the font size of each OSN name  is relative to the network's number of active users~\cite{100millionactiveusers}.}
\label{fig:cloud}

\end{figure}

In recent years, global online social network (OSN) usage has increased sharply as these networks \s{have} become interwoven into people's everyday lives  as \s{virtual  meeting} places \s{that facilitate communication}.  \s{OSNs}, such as Facebook~\cite{facebook}, Google+~\cite{googleplus}, LinkedIn~\cite{linkedin}, Sina Weibo~\cite{weibo}, Twitter~\cite{twitter}, Tumblr~\cite{tumblr}, and VKontakte (VK)~\cite{vk}  have hundreds of millions of daily active users (see \s{Fig.}~\ref{fig:cloud}). 
Facebook, for example, has more than \s{1.23} billion monthly active  users, \s{945} million of which are active mobile Facebook users as of \s{December} 2013~\cite{fbstats2013}. 

\s{Facebook users have a total of over 150 billion friend connections and upload \s{on average more than 350 million photos} to Facebook each day~\cite{fb2012_report}}. Unfortunately, many OSN users are unaware of the security risks which exist in these types of communications, including privacy risks~\cite{boshmaf,mislove}, identity theft~\cite{bilge}, malware~\cite{baltazar}, fake profiles (also in some cases referred to as sybils~\cite{cao2012aiding, stringhini2013follow} or socialbots~\cite{boshmaf,boshmaf2012design,elishar}), and sexual harassment~\cite{wolak,ybarra2008risky}, among others.  
\s{A study by Dwyer et al.~\cite{DwyHilPas07} found that Facebook and MySpace~\cite{myspace} users trust these OSNs, and  they have trust in other users within these social networks. This trust leads to information sharing and  to developing new relationships. }
Moreover, according to recent studies~\cite{acquisti,boshmaf}, many OSN users expose personal and intimate details about themselves,  their friends, and their \s{relationships}, whether by posting photos or by directly providing information such as a home address and a phone number. 
Furthermore, according to Boshmaf et al.~\cite{boshmaf} and Elyashar et al.~\cite{elishar,elishar2}, Facebook users have been shown to accept friendship requests from people whom they do not know but \s{with  whom they} simply have several friends in common. By accepting these friend requests, users unknowingly disclose their private information to total strangers. 
This information could be used maliciously, harming users both in the virtual and in the real world. These risks escalate when the users are \s{young} children or teenagers who are by nature more exposed and vulnerable than adults. 

As the use of \s{OSNs} becomes progressively more embedded \s{in users' daily lives}, personal information becomes easily exposed and abused.
Information harvesting, by both the \s{OSN} operator itself and by third-party commercial companies, has recently been identified as a significant security concern for OSN users. Companies can \s{exploit} the harvested personal information for a variety of purposes, all of which can jeopardize a user's privacy. For example, companies can use collected private information to tailor online ads according to a user's profile~\cite{tucker2011social}, to gain profitable insights about their customers, or even to share the user's private and personal data with the government~\cite{prism}. This information may include general data, such as age, gender, and income; however, in some cases more \s{delicate and potentially harmful} information can be exposed, such as the user's sexual orientation~\cite{jernigan2009gaydar} and if the user \s{has} consumed addictive substances~\cite{kosinski2013private}. These privacy \s{concerns} become more alarming when considering the nature of OSNs: information \s{regarding a network user} can be obtained without even directly accessing \s{the} individual's online profile; personal details can be inferred solely by collecting data on the user's friends~\cite{mislove}.

To cope with the above-mentioned threats, multiple solutions have been offered by OSN operators, security companies, and academic researchers. \s{OSNs}, like Facebook, attempt to protect their users by adding authentication processes to ensure that the registered user is a real person~\cite{boshmaf,facebookid,facebooktag,twitter2factors,facebook2factor}.
Moreover, many OSN operators also support a configurable user privacy setting that enables users to protect their personal data from other users within the network~\cite{liu2011analyzing,mahmood2011poster}. As for privacy settings, \s{OSN} operators currently face a conflict of interest: On the one hand, \s{since} personal information is a commodity, the more that is shared, the better.  On the other hand, a user who is anxious about \s{his or her} privacy is a liability and will probably share less information and become \s{consequently}  less active. Nevertheless, \s{both} regulating authorities and public groups try to \s{address} privacy concerns \s{and make them a part of } public discourse and consideration~\cite{sale}. Today there are additional protection mechanisms which include defenses against spammers~\cite{aggarwal2013detection,benevenuto2009detecting,bhatcommunity,debarr,lee,stringhini,wang}, fake profiles~\cite{cao2012aiding,danezis2009sybilinfer,privacyprotector,fire2013friend,fire2012strangers, tran2009sybil,wang2013you,wang2012social,yu2008sybillimit,yu2006sybilguard}, and other threats. For example, security companies like Check Point~\cite{checkpoint}, Websense~\cite{websense}, and \s{Infoglide~\cite{minormonitor}} offer social tools to protect users in the OSN world. These companies typically offer products which monitor user activity in order to identify and protect users. The modern day threats are so pervasive that even the academic community has addressed this issue by publishing studies which attempt to solve different OSN threats and offer improvements in identity protection~\cite{debarr,  privacyprotector, fire2012strangers, kontaxis,lee,wang}.

\s{
\subsection{Contributions}
This paper presents the ``big picture'' of the current state-of-the-art academic and industry solutions that can protect OSN users from various security and privacy threats. More specifically, this study offers the following contributions: First, we outline the OSN threats that target every user of social networks, with an additional focus on young children and teenagers. 
Second, we present a thorough overview of the existing solutions to these threats, namely those provided by OSN operators, commercial companies, and academic researchers. Third, we compare and discuss the protection ability of the various solutions. Lastly, we give easy-to-implement recommendations on how OSN users can better protect their security and privacy when using social networks.
 }

\subsection{Organization} 

The remainder of the paper is organized as follows: In Section~\ref{osnusage}, we introduce insightful  statistics  on OSNs usage. Next, in Section~\ref{sec:threats}, we describe different types of OSN threats. Section~\ref{sec:solutions} follows with various solutions to assist in protecting \s{social network} users. \s{In Section~\ref{sec:disc}, we discuss the various presented threats and their corresponding solutions.} In Section~\ref{sec:recom}, we offer recommendations that OSN users \s{can apply} in order to improve their online security and privacy. \s{Next, in Section~\ref{sec:future}, we offer future research directions. Our conclusions are presented in Section~\ref{sec:conclousions}.}   

\section{Online Social Network Usage}
\label{osnusage}
\s{Today many OSNs} have tens of millions \s{of} registered users. Facebook, with more than \s{a} billion active users, is currently the largest and most popular OSN in the world~\cite{facebook-newsroom}. Other well-known OSNs are Google+, with over 235 million active users~\cite{googleplusstat}; Twitter, with over 200 million active users~\cite{twitterstat}; and LinkedIn, with more than 160 million active users~\cite{linkedinpress}. While some experts insist that \s{OSNs} are a passing fashion and will eventually be replaced by another Internet fad, current user statistics concur that OSNs are here to stay. A recent survey by the Pew Research Center's Internet and American Life Project~\cite{pew2013} revealed that 72\% of online American adults use social networking sites, a dramatic increase from the 2005 Pew survey which discovered that just 8\% of online American adults used social networking sites. Moreover, the survey revealed that 89\% of online American adults between the ages of 18 to 29 use social network sites, while in 2005 only 9\% of the survey participants  in this age group used this type of site. These survey results are compatible with a previous report published by Nielsen in 2011~\cite{nielsen}, disclosing that Americans spent 22.5\% of their online time on OSNs and blogs, more than twice the time spent on online games (9.8\%). Other common activities that consume Americans' \s{online} time include email (7.6\%); portals (4.5\%); videos and movies (4.4\%); \s{searches (4.0\%), and instant messaging (3.3\%)}. The amount of collective time spent on OSNs, especially on Facebook, is enormous and ever-growing.  U.S. users spent a total of 53.5 billion minutes on Facebook during May 2011, 17.2 billion minutes on Yahoo~\cite{yahoo}, and 12.5 billion minutes on Google~\cite{google}. 

Mobile devices, or cellular phones, \s{increasingly} serve as platforms for Internet usage.
\s{According to Facebook's report~\cite{fbstats2013} in December 2013, Facebook had 556 million daily active mobile users, an increase of 49\% year over year. 
Additionally, Facebook and Google+ mobile applications are the second and fourth (respectively)  most frequently used smartphone applications~\cite{mobileapp_freq}.} It  should be noted that the use of OSNs on mobile devices not only \s{promotes} an even ``closer \s{relationship}'' to social networks but also  \s{can} pose additional privacy concerns, especially \s{regarding the  collection} of location data and \s{the opportunity} for advertisers to identify specific types of users.          	   

Besides being popular among adults, \s{OSNs} have become \s{extremely popular with young} children and teenagers. A comprehensive study~\cite{livingstone2011eu} carried out in 25 European countries with 25,000 participants produced the following statistics: 60\% of children \s{9 to 16 years old} who access the Internet use it daily (88 minutes of use on average) and 59\% of \s{ those 9 to 16 years old} who use the Internet have a personal OSN site profile (26\% of \s{ages 9 to 10}; 49\% of \s{ages 11 to 12}; 73\% of \s{ages 13 to 14}; 82\% of \s{ages 15 to 16}). Note that the terms of use governing  OSNs do not officially allow users under the age of 13.  
Furthermore, 26\% of the children in \s{this same} European study had their social network profile set to ``public'' (\s{i.e.}, accessible to strangers), 14\% reported having their address or phone number listed on their profile, and 16\% admitted that their profile displayed an inaccurate age.   
\s{In addition,} 30\% of the children surveyed reported having an online connection with a person they had never met face to face, 9\% reported having actually met face to face with someone with whom they had only an online connection, 9\% reported experiencing a misuse of personal data, 21\% reported encountering one or more types of potentially harmful user-generated content, and 6\% reported receiving  malicious or hurtful messages on the Internet\s{~\cite{livingstone2011eu}}.
These findings reiterate our previous claim: the use of OSNs is embedded in the everyday lives of children and teenagers, and can result in personal information being exposed, misused, and potentially abused. Interestingly, about a third of the parents in this European study claimed that they filter their children's use of the Internet, while a quarter specifically stated that they use monitoring tools\s{~\cite{livingstone2011eu}}.

\section{Threats}
\label{sec:threats}
With the increasing usage of \s{OSNs}, many users have unknowingly become exposed to threats both \s{to} their privacy and \s{to} their security. These threats can be divided into \s{four} main categories. The first category \s{contains} \textit{classic threats}, namely, \s{privacy and security} threats that not only jeopardize OSN users but \s{also} Internet users not using social networks (see Section~\ref{sec:classical}).  The second category \s{covers}  \textit{modern threats}, \s{that is}, threats that are mostly unique to the environment of \s{OSNs} and which use the OSN infrastructure to endanger user privacy and security (see Section~\ref{sec:modern}).  
\s{The third category consists of \textit{combination threats}, where we describe how today's attackers can, and often do, combine various types of attacks in order to create more sophisticated and lethal attacks (see Section~\ref{sec:combination})}.
\s{The fourth and last category includes threats specifically  targeting children who use social networks 
(see Section~\ref{sec:children}).} 

\begin{figure}[hb]
\begin{center}
\includegraphics[
width=\textwidth]{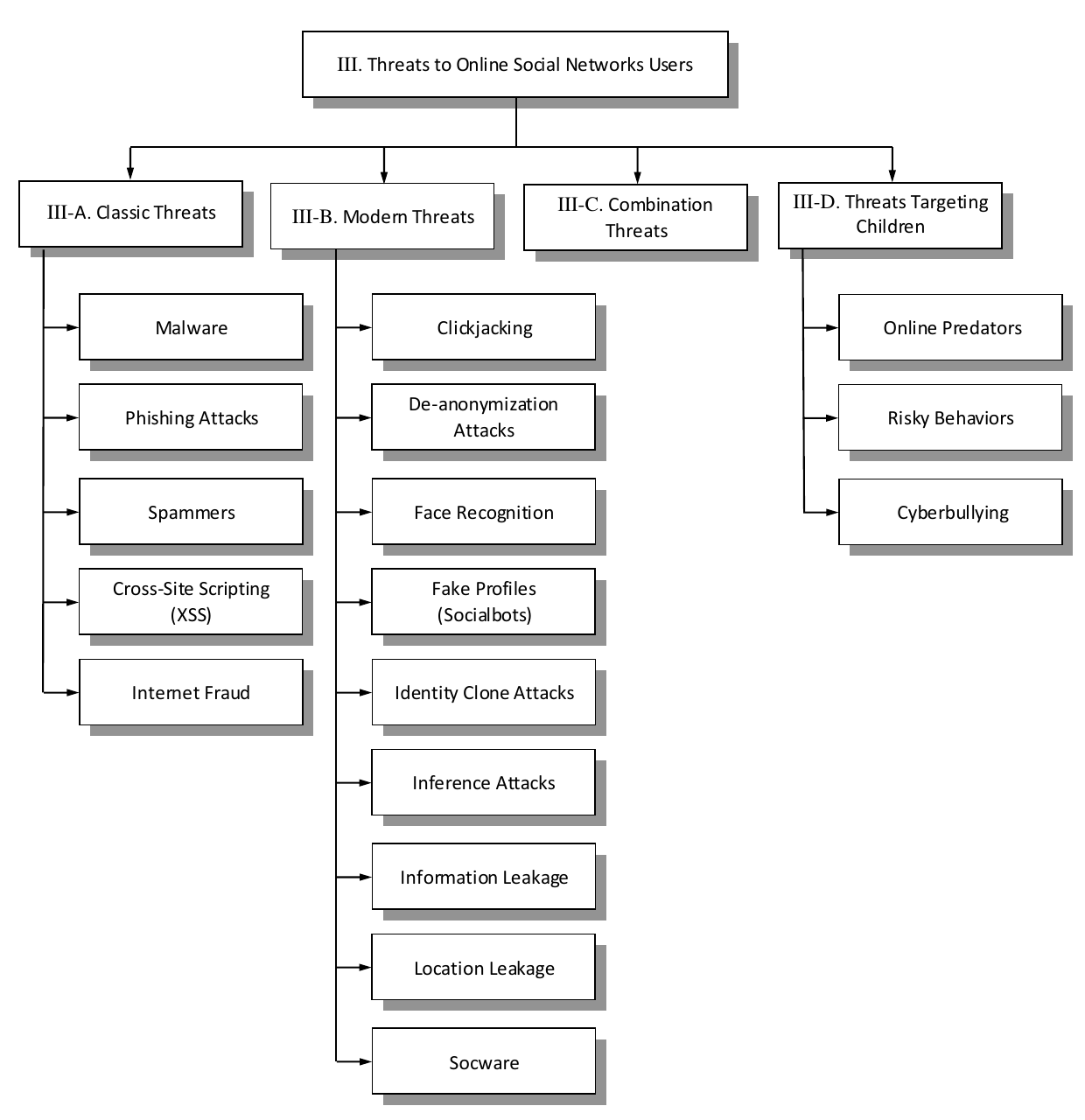}
\end{center}
\caption{ \textbf{Threats to Online Social Network Users.}}
\label{fig:threats}
\end{figure}

\s{ Fig.~\ref{fig:threats} diagrams all the specific threats listed in the following sections. The boundaries between all these categories of threats, however, can become blurred as techniques and targets often overlap.} 

\subsection{Classic Threats}
\label{sec:classical}
Classic threats have been a problem ever since the Internet gained
widespread usage. Often referred to as malware, \s{spam}, cross-site scripting (XSS) attacks, or phishing, they continue to be an ongoing issue.
Though these \s{threats} have been addressed in the past, they have become \s{increasingly} viral due to the structure and nature of OSNs and \s{can} spread quickly among network users. 
Classic threats can take advantage of a user's personal information published in a social network to attack not only the user but also their friends simply by adjusting the threat to accommodate the user's personal information. 

For example, an attacker can plant a malicious code inside an attractive spam message that employs a user's details from his or her Facebook profile. Due to the personal nature of this crafted message, the chances that the innocent user will open the message and get infected are likely. In many cases, these threats target essential and everyday user resources such as credit card numbers, account passwords, computing power, and even computer bandwidth (in order to send spam emails). \s{Alarmingly, these types of threats can also exploit the infected user's stolen credentials to post messages on the user's behalf or even change the user's personal information. }

The different classic threats are \s{described} below, along with real-life scenarios where these types of menaces have jeopardized a real user's privacy and security.

\paragraph{Malware.} Malware is malicious software developed to disrupt a computer operation in order to collect a user's credentials and gain access to his or her private information. Malware in \s{social networks} uses the OSN structure to \s{propagate} itself \s{among} users and their friends in the network. \s{In some cases, the malware can use the obtained credentials \s{to impersonate }the user and send contagious messages to the user's online friends}. Koobface was the first malware to successfully propagate through \s{OSNs such as} Facebook, MySpace, and Twitter. Upon infection, Koobface attempts to collect login information and join the infected computer in order to be part of a botnet~\cite{baltazar}, a so-called ``zombie army'' of computers which \s{often} is then used for criminal activities, such as sending spam messages and attacking other computers and servers \s{over} the Internet.

\paragraph{Phishing Attacks.} Phishing attacks are a form of social engineering to acquire user-sensitive and private information by impersonating a trustworthy third party. A recent study~\cite{amin2010facebook} showed that users \s{who interact} on social networking \s{websites} are more likely to fall for phishing scams due to their \s{social and trusting nature}. Moreover, in recent years, phishing attempts within OSNs have increased sharply. According to the Microsoft Security Intelligence Report~\cite{microsoft-report}, 84.5\% of all phishing attacks target social \s{network site users}. One such phishing attack occurred \s{on Facebook,  luring} users onto fake Facebook login pages. Then, the phishing attack spread among Facebook users by \s{inviting friends} to click on a link posted on the original user's \s{profile space}~\cite{mills2009facebook}. Fortunately, Facebook acted to stop this attack.

\paragraph{Spammers.} Spammers are users who use electronic messaging systems in order to send unwanted messages, like advertisements,  to other users. \s{OSN} spammers use the \s{social networking platform to send} advertisement messages to other users by creating fake profiles~\cite{fire2012strangers}. The spammers can also use the OSN platform to add comment messages to pages which are viewed by many users in the network.  
An example of the prevalence of network spamming can be found \s{on} Twitter, which has suffered from a massive amount of spam. In August 2009, 11\% of Twitter messages were spam messages. However, by the beginning of 2010, Twitter had successfully cut down the percentage of spam message to 1\%~\cite{twitterspam}. \s{However, a 2013 article~\cite{twitterspam2013} states ``Social spam, as it already exists on Twitter, will continue to grow and unless the company addresses the problem quickly, it may be the one thing that sinks it.''}

\paragraph{Cross-Site Scripting (XSS).} \s{An XSS} attack is an assault against web applications. The attacker who uses the \s{XSS} exploits the trust of the web client in the web application and causes the web client to run malicious code capable of collecting sensitive information. OSNs, which are types of  applications, can suffer from XSS attacks. Furthermore, attackers can use an XSS vulnerability combined with the OSN infrastructure to create an XSS worm that can spread virally among social network users~\cite{livshits2008spectator}. In April 2009, such an XSS worm, called Mikeyy, rapidly transmitted automated tweets across Twitter and infected many users, among them celebrities like Oprah Winfrey and Ashton Kutcher. The Mikeyy worm used an XSS weakness and the Twitter network structure to spread through Twitter user profiles~\cite{mikeyy}.

\paragraph{Internet Fraud.} Internet fraud, also known as cyber fraud, refers to using Internet access to scam or take advantage of people. In the past, con artists used traditional in-person social networks, such as weekly group meetings, to gradually  establish strong bonds with their potential victims.  \s{Currently}, according to the North American Securities Administrators Association (NASAA)~\cite{nasaa}, with the rising popularity of \s{online networking}, con artists have turned to OSNs to establish trust connections with their victims, and \s{then they take advantage of}  personal data published in the victims' online profiles.  In recent years, for example,  fraudsters have been hacking into the accounts of Facebook users  who travel abroad. Once they manage to log into a user's account, the scammers cunningly  ask the user's friends for assistance in transferring money to the scammer's bank account. One victim of this type of fraud was Abigail Pickett. While travelling in Colombia, Abigail discovered that her Facebook account had been hijacked \s{by} someone in Nigeria\s{, and  it was} being used to send requests for money \s{to her network} friends on the pretext that she was ``stranded''~\cite{fraud}.

\subsection{Modern Threats}
\label{sec:modern}

Modern threats \s{are typically} unique to \s{OSN environments.}
Usually these threats specifically target users' personal information as well as the personal information of their friends. 
\s{For example, an attacker who is trying to gain access to a Facebook user's high school name \textemdash viewable only by the user's Facebook friends \textemdash can  create a fake profile with pertinent details and initiate a friend request to the targeted user. If the user accepts the friend request, his or her details will be exposed to the attacker. Alternatively, the attacker can collect data from the user's Facebook friends and employ an inference attack to infer the high school name from the data collected from the user's friends.}

\s{In what follows}, we illustrate the various modern threats and real-life scenarios where these types of threats have jeopardized an OSN user's privacy and security.

\paragraph{Clickjacking.} Clickjacking is a malicious technique which tricks users into clicking on something different from  what they intended to click. By using clickjacking, the attacker can manipulate the user into posting spam messages on his or her Facebook timeline, performing ``likes'' to links unknowingly (also referred as likejacking), and even opening  a microphone and web camera to record the user~\cite{lundeennew}. An example of a clickjacking attack occurred \s{on} Twitter in 2009 when Twitter was plagued by a ``Don't Click'' attack. The attacker tweeted a link with the message ``Don't Click'' along with a masked URL (the actual URL domain was hidden). When Twitter users clicked on the ``Don't Click'' message, the message automatically spread virally and was posted onto their Twitter accounts~\cite{dontclick}.

\paragraph{De-Anonymization Attacks.} In many \s{OSNs} like Twitter and MySpace, users can protect their privacy and anonymity by using pseudonyms. De-anonymization attacks use techniques such as \s{tracking cookies}, network topology, and user group memberships to uncover the user's real identity. An example of de-anonymization  was demonstrated 
\s{ by Krishnamurthy and Wills~\cite{krishnamurthy2009leakage}, who proved that it is possible for third parties to uncover OSN user identities by linking information leaked via social networking sites. Krishnamurthy and Wills also showed that most users on the studied OSNs were vulnerable to having their OSN identity information leaked via tracking mechanisms, such as tracking cookies.
Another example of this type of attack} was presented by Wondracek et al.~\cite{wondracek}; they offered a method to de-anonymize users in OSNs by using only the users' group memberships. Wondracek et al. tested their method on the Xing~\cite{xing} OSN and succeeded in identifying 42\% of the users. 
An additional recent example was presented by Peled et al.~\cite{peled2013}, who introduced a method for matching user profiles across several OSNs. The method was evaluated by matching profiles across  Facebook and \s{Xing.}

\s{
\paragraph{Face Recognition.} Many people use OSNs for uploading pictures of themselves and their friends. 
Millions and millions of photos are uploaded to Facebook each day~\cite{fb2012_report}.
 Moreover, many Facebook user profile pictures are publicly available to view and download. For instance, the Faces of Facebook website~\cite{fbfaces} allows Internet users to view the profile images of over 1.2 billion Facebook users. These photos can be used to create a biometric database, which can then be used to identify OSN users without their consent.

In 2011, Acquisti et al.~\cite{acquisti2011faces} demonstrated the threat of face recognition to OSN user privacy by performing three experiments. The first experiment showed that it is possible to match  ``online to online'' image datasets by using publicly accessible Facebook user profile pictures to re-identify profiles on one of the most popular dating sites in the United States. In their second experiment, Acquisti et al. demonstrated that ``offline to online'' image datasets can also be matched. Namely, they used publicly available images from Facebook to identify students strolling  through campus. In their third experiment,  
Acquisti et al. illustrated that it is possible to predict personal and sensitive information from a face;
an individual's interests, activities, and even his or her social security number could be automatically predicted by matching the face image with the person's Facebook image to obtain the person's full name. Following this action, the attacker could use the obtained name to cross-reference it against other datasets. 
}

\paragraph{Fake Profiles.} Fake profiles (also referred to as sybils or socialbots) are automatic or semi-automatic profiles that mimic human behaviors in \s{OSNs}. In many cases, fake profiles can be used to harvest \s{users'} personal data from social networks. By initiating friend requests to other users in the OSN, who \s{often} accept the requests, the socialbots can gather a user's private data which should be exposed only to the user's friends. 
Moreover, fake profiles can be used to initiate sybil attacks~\cite{douceur2002sybil}, publish spam messages~\cite{gao2010detecting}, or even manipulate OSN statistics~\cite{stringhini2013follow,limitedrun}. A recent article asserted that the market of buying fake followers and fake retweets is already a multimillion-dollar business~\cite{twitterny}.
Additional approaches that \s{generate} fake profiles were demonstrated recently by Boshmaf et al.~\cite{boshmaf} \s{when  
an} army of more than a hundred Facebook socialbots was created, which then attempted to infiltrate innocent Facebook profiles by initiating a series of friend requests. The socialbot army succeeded in generating approximately 250GB of inbound \s{Facebook traffic}.
Moreover, the socialbot friend acceptance rates climbed to 80\% whenever a socialbot and an innocent Facebook user had more than eleven friends in common. In some cases, even one well-manipulated fake profile can cause \s{extensive} damage as proven by Thomas Ryan, who assumed the fictional profile of Robin Sage to connect to hundreds of users from various social networking sites~\cite{ryan2010getting}. 

\paragraph{Identity Clone Attacks.} Using this technique, attackers duplicate a user's online presence \s{either} in the same network, or across different networks, \s{to} deceive the cloned user's friends into forming a trusting relationship with the cloned profile. The attacker can use this trust to collect personal information about the user's friends or to perform various types of online fraud. 
\s{An example of an identity clone attack} occurred recently with NATO's most senior commander, Admiral James Stavridis. His profile details were cloned and \s{then} used to collect data on defense ministry officials and other government officials by tricking them into becoming  friends with the newly cloned Facebook profile~\cite{clone}.

\paragraph{Inference Attacks.} Inference attacks in \s{OSNs} are used to predict a user's personal, sensitive information that the user has not chosen to disclose, such as religious affiliation \s{or} sexual orientation. These \s{types} of attacks can be implemented using  data mining techniques combined with publicly available OSN data, such as network topology and \s{data from} users' friends. 
An inference attack was demonstrated by Mislove et al.~\cite{mislove} who presented techniques for predicting a user's attributes based on other users' attributes in the OSN. They tested their techniques and inferred different Facebook users' attributes, such as educational information, personal preferences, and geographic information.
Recently, \s{inference} attacks on organizations \s{were explored} by Fire et al.~\cite{orgmining}. They presented an algorithm for inferring the OSN of a targeted organization based \s{solely} on publicly available data from \s{social networks}. Fire et al. tested their algorithm on six organizations of different scales \s{using} publicly available data from the Facebook profiles of the organization's employees, resulting in a successful reconstruction of the social networks \s{within} these six organizations. Additionally, certain details \s{ could be inferred about the targeted organizations,} some of \s{which were confidential.}

\s{
\paragraph{Information Leakage.} OSNs allow users to openly share and exchange information with their friends and other users in the network. In some cases OSN users willingly share sensitive information about themselves and other people, such as \s{health-related} information~\cite{mao2011loose,sadegh2013hri} and sobriety status~\cite{mao2011loose}. In a recent study, Torabi and Beznosov~\cite{sadegh2013hri} observed that 95.8\% of 166 participants shared some health-related information through their OSN accounts. Leakage of sensitive and personal information may have negative implications for the social networks users. For example, insurance companies may use OSN data to  identify risky clients~\cite{insurence2010}. These companies can use OSN leaked information  to detect clients with medical conditions, consequently increasing their premiums or denying their coverage. 
Additionally, employers use social networks for screening job  applicants~\cite{vicknair2010use}. Therefore, leaking personal information, such as drinking habits, on OSNs may jeopardize future chances for finding employment. 
}

\s{
\paragraph{Location Leakage.\footnote{\s{Location leakage is a private case of information leakage, which was discussed in the previous paragraph. However, due to serious privacy threats that could occur as a result of location leakage, such as location monitoring and stalking, we present this threat in a separate subsection.}}} With the increasing use of smart mobile devices that encourage sharing of location information~\cite{humphreys2007mobile}, many people use OSNs to willingly share private and sometimes sensitive information about their (or their friends') current or future whereabouts. A study by Humphreys et al.~\cite{humphreys2010much} found that 20.1\% of examined Twitter tweets included information on when people were engaging in certain activities, and 12.1\% of the tweets mentioned the person's location.
Additionally, a study by Mao et al.~\cite{mao2011loose} demonstrated that classifiers can be trained to identify Twitter users' locations in real time. 
Moreover, Cheng et al.~\cite{Cheng:2010:YYT:1871437.1871535} presented a framework for estimating a user's city-level location based on the content of the user's tweets.
This type of information can be used by criminals and stalkers. 
For example, Israel Hyman from Arizona tweeted that he was looking forward to his family vacation to St. Louis. He also tweeted
again once he had arrived in Missouri. When Hyman returned home, he discovered that his house had been burglarized~\cite{humphreys2010much}.
An even more disturbing example of location leakage threats is given by the website Pleaserobme.com~\cite{robme,grove2010rob}, which  a way to find the location information of specific Twitter and Foursquare~\cite{foursquare} users. 

In some cases, OSN users unknowingly share their locations by uploading media items, such as photos and videos, which may be embedded with geotagging information about their current and past locations~\cite{Friedland:2010:CJP:1924931.1924933}. For example, Adam Savage, the host of the popular science program \textit{MythBusters}, posted a picture on Twitter of his car parked in front of his house. The uploaded image contained a geotag which exposed the place where the photo was taken~\cite{photos2010ny}.   
}
\paragraph{Socware.} Socware  entails fake and possibly damaging posts and messages from friends in \s{OSNs}. 
Socware may lure victims by offering false rewards to users who install 
socware-related malicious Facebook applications or visit questionable socware \s{websites}.
After the users have cruised the socware website or installed the relevant application, \s{the installed socware sends messages on the user's behalf to the user's friends}, essentially assisting the socware viral spread~\cite{rahman2012efficient}.
In 2012, Rahman et al.~\cite{rahman2012efficient} investigated over 40 million posts and discovered that 49\% of  the studied  users were exposed to at least one socware post in a four-month period. Moreover, Rahman et al.~\cite{rahman2012frappe} discovered that 13\% of 111,000 studied applications were malicious applications that could assist in spreading socware. Additionally, a recent study by Huang et al.~\cite{huang2013analysis} studied the ecosystem which enables socware to propagate (cascade). By analyzing data from the \s{profile pages} of approximately 3 million Facebook users over a period of five months, they discovered that ``socware cascades are supported by Facebook applications that are strategically collaborating with each other in large groups.'' 

\s{
\subsection{Combination Threats}
\label{sec:combination}
Today's attackers can also combine classic and modern threats in order to create a more sophisticated attack. For example, an attacker can use a \s{phishing attack} to collect \s{a targeted} user's Facebook password and then post a message \s{containing} a clickjacking attack on the \s{targeted} user's timeline, \s{thus luring the user's} Facebook friends to click on the posted message and install a hidden virus onto their own computers. Another example is \s{the use of} cloned profiles to collect personal information about \s{friends of the cloned user}. Using the friends' personal information, the attacker can send uniquely \s{tailored} spam email \s{messages} containing a virus. By using personal information, the virus is more likely to be activated. 

Note that the recovery processes \s{from classic and modern} threats are distinct.  In order to recover from a classic attack, like a virus, it is usually possible to simply reinstall the operating  system, change the current passwords, and cancel the affected credit cards. However, in order to recover from a modern OSN attack that ``steals your reality''~\cite{altshuler2011stealing}, more effort must be \s{made} because resetting personal information is excessively time consuming and not always possible. For instance, you \s{could} change your email address, \s{but it would be much more difficult to change your home address.}} 

\subsection{Threats Targeting Children}
\label{sec:children}
\s{Children, whether young children or teenagers, certainly experience the classic and modern threats detailed above, but there are also threats that intentionally and specifically target younger users of OSNs. Due to the critical nature of this topic, this section highlights those threats, as well as describes specific findings from current studies. }

\paragraph{Online Predators.} The greatest concern \s{regarding} the personal information safety of children \s{relates to} Internet pedophiles, also referred to as online predators. Livingstone and Haddon~\cite{childsaftey} of EU Kids Online defined a typology in order to understand the risk and harm related to the following online activities: harm from \textit{content} (\s{a} child's exposure to pornography or harmful sexual content), harm from \textit{contact} (a child who is contacted by an adult or another child for the \s{purpose} of sexual abuse), and harm from \textit{conduct} (the child as an active initiator of abusive or risky behaviors). 
Behaviors \s{that are} considered to be Internet sexual exploitation of children include adults using children for the production of child pornography and its distribution, consumption of child porn, and the use of the Internet as a means to initiate online or offline sexual exploitation. 
In their study from 2008, Wolak et al.~\cite{wolak} critically examined the myth and reality of the online predator. The image of \s{an} Internet predator in the media is \s{that} of an adult man
\s{who pretends to be a friend to an innocent young boy or girl through whom he 
collects} personal data; \s{he hides} his sexual intentions until the actual meeting, which likely involves rape or kidnapping. According to Wolak et al., \s{however,} the truth is far more complex. Wolak et al. assert that most Internet-initiated sex crimes indeed start with establishing a relationship between an adult and a child through the use of instant messaging, emails, chats, etc. However, in most cases children are aware of the fact that they are talking to an adult, and if the relationship escalates to attending a real-life meeting, they are aware and to some extent expect to engage in sexual activity. More often than not, the encounter involves non-forcible sexual activity, yet it is with a person under the age of consent and therefore constitutes a crime.  Contrary to the common notion, \s{Wolak et al. discovered that} most victims of Internet-initiated sex crimes were teenagers (aged \s{13 to 17}), and none under age 12 were reported~\cite{wolak}. Therefore, these crimes do not constitute the clinical definition of pedophilia: ``the fantasy or act of sexual activity with prepubescent children'' \s{~\cite{pedodef}}. \s{Of course, this does not make the crimes any less distasteful.}

\paragraph{Risky Behaviors.} Potential risky behaviors \s{of children} may include direct online communication with strangers, use of chat rooms for interactions with strangers, sexually explicit talk with strangers, {and} giving private information and photos to strangers. It should be noted that while each of the above-mentioned behaviors alone poses a risk, the combination of a few of these behaviors can justifiably cause  enormous anxiety regarding a child's safety.
\s{Wolak et al.~\cite{wolak} maintain that risky online behaviors and specific populations who are more exposed to them can be identified.}
\s{Additionally,} there is a well-established link between online and offline behaviors. Researchers contend that victims of Internet abuse are very often vulnerable children, such as youths with a history of physical or sexual abuse or those who suffer from depression or social interaction problems\s{~\cite{wolak}}. All children living with these kinds of issues are at a higher risk of sexual abuse on the Internet or through \s{online-initiated} encounters\s{~\cite{wolak}}. 

\paragraph{Cyberbullying.} Cyberbullying (also referred to as cyber abuse) is bullying that takes place within technological communication platforms, such as emails, chats, phones conversations, and OSNs, by an attacker who uses the platform to harass his victim by sending repeated hurtful messages, sexual remarks, or threats; by publishing embarrassing pictures or videos of the victim; or by engaging in other inappropriate behavior. Today, cyberbullying has become  \s{a common phenomenon} in \s{OSNs} in which the attacker can utilize the \s{network's} infrastructure to spread cruel rumors about the victim and share embarrassing pictures with the victim's network of friends~\cite{deans2012story}. Cyberbullying usually affects children, \s{rather than adults}. A recent online survey, which included 18,687 parents from 24 countries,  \s{revealed}  that 12\% of parents claim their child has been cyberbullied~\cite{lpsos}. Additionally, according to the survey's results, \s{the} majority of children experienced this harassing behavior on widely used social networking sites like Facebook. 
Horrifically, in some cases cyberbullying can \s{cause} catastrophic  results, as in the cases of Amanda Michelle Todd\s{~\cite{deans2012story}} and Rebecca Ann Sedwick~\cite{rebecca}, both of whom committed suicide after being cyberbullied on Facebook.

\section{Solutions}
\label{sec:solutions}
\begin{figure}[H]
\begin{center}
\includegraphics[
width=\textwidth]{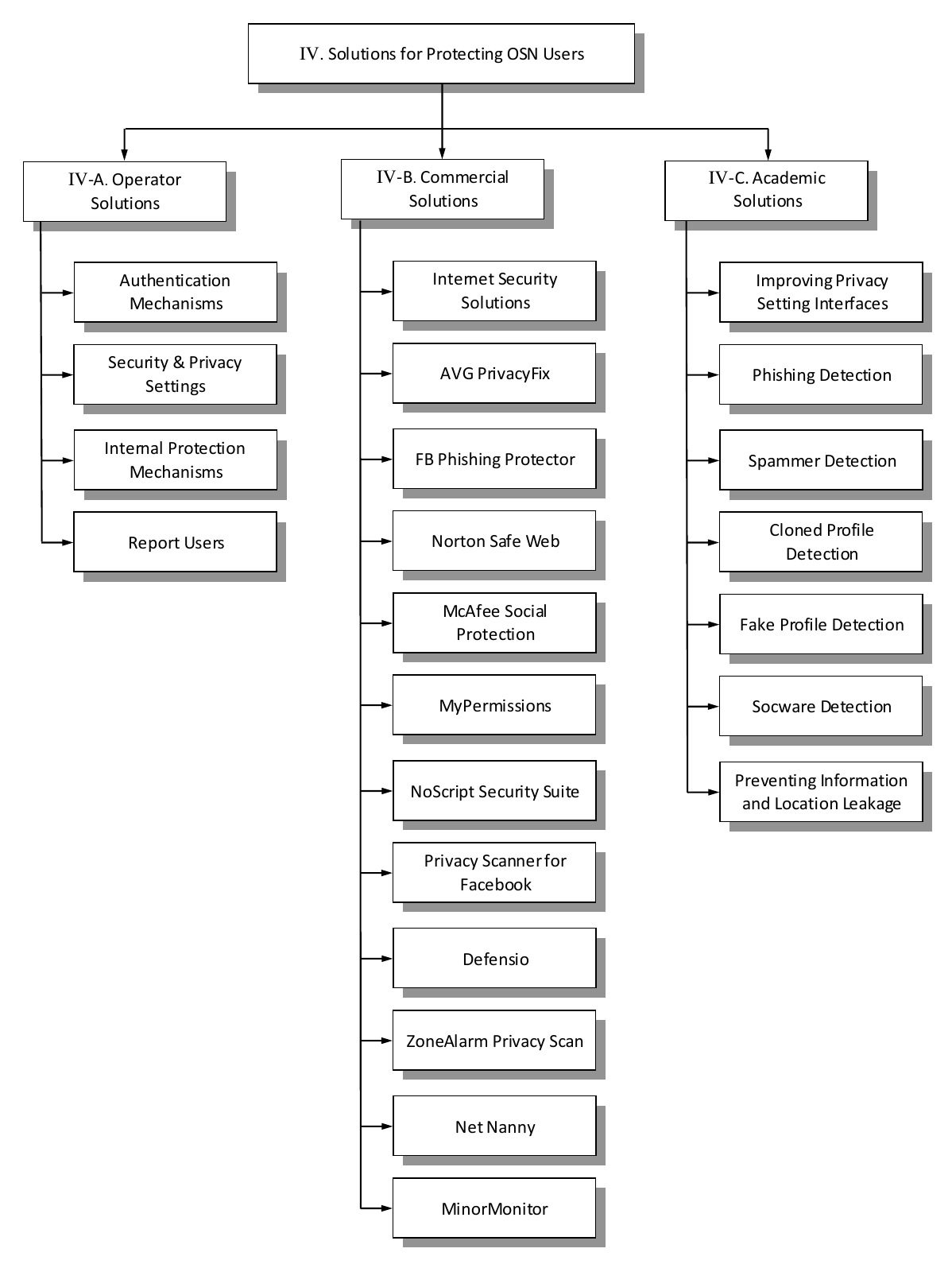}
\end{center}
\caption{ \textbf{Security and Privacy Solutions for Online Social Networks.}}
\label{fig:solutions}
\end{figure}

In recent years, social network operators, security companies, and academic researchers have tried to deal with the above-mentioned threats \s{by} proposing a variety of solutions (see Fig.~\ref{fig:solutions} and Table~\ref{tab:threats_and_solutions}). In this section we describe possible solutions which can assist in protecting the security and privacy of \s{OSN} users.

\subsection{Social Network Operator Solutions}
\label{sec:operator}
\s{OSN} operators attempt to protect their users by activating safety measures, such as employing user authentication mechanisms and applying user privacy settings. Several of these techniques are described in detail below.

\paragraph{Authentication Mechanisms.} In order to make sure the user registering or logging into the social network is a real person and not a socialbot or a compromised user account, OSN operators \s{use} authentication mechanisms, such as CAPTCHA~\cite{boshmaf},  photos-of-friends identification~\cite{facebooktag}, multi-factor authentication~\cite{facebook2factor},  and in some cases even requesting that the user send a copy of his or her government issued ID~\cite{facebookid}. As an example, Twitter recently introduced its two-factor authentication mechanism~\cite{twitter2factors}, requiring the user to not only insert a password when logging into Twitter but also provide a verification code that was sent to the user's mobile device.  

This mechanism prevents a malicious user from logging in through hijacked accounts and publishing false information through those hijacked accounts. Such a mechanism would thwart incidents such as when hackers hijacked the Associated Press (AP) Twitter account, resulting in the rapid propagation of false information about explosions in the White House, which caused panic on Wall Street~\cite{twitterap}.

\paragraph{Security and Privacy Settings.} Many \s{OSN}s support \s{various} configurable user privacy \s{settings} that \s{enable} users to protect their personal data from other users \s{or applications~\cite{liu2011analyzing,mahmood}.}  
\s{ Facebook users, for example, can customize their privacy settings and choose which other users in the network (such as Friends, Friends of Friends, and Everyone) are able to view their details, pictures, posts, and other personal information~\cite{facebookprivacy}.  
A similar example of customizable privacy settings exists in Google+: users place each one of their friends into \s{groups, also known as circles, such as Best Friends circle, Work circle, and High School Friends circle.} Using these circles, Google+ users can better protect their privacy by deliberately choosing which of their posts are exposed to each circle~\cite{googlep2011}.
Moreover, both Facebook and Google+ enable their users to approve or revoke the access of applications to the users' personal data~\cite{facebook_revoke,googlep_revoke}.}

Some \s{OSNs} also support extra security configurations which enable the user to activate secure browsing, receive login notifications, and establish other safety features~\cite{securitysettings}.
\s{However, many \s{OSN} users \s{still} simply maintain the default privacy settings, letting their data be exposed to strangers~\s{\cite{fire2013friend,krishnamurthy2009leakage}}.}

\paragraph{Internal Protection Mechanisms.} Several \s{OSNs} protect their users by implementing additional internal protection mechanisms for defense against spammers, fake profiles, scams, and other threats~\cite{twitterspam,stein_2011facebook}. Facebook, for example,  protects its users from malicious attacks and information collecting by activating the Facebook Immune System (FIS). The FIS is described as an adversarial learning system that performs real-time checks and classifications on \s{read-and-write} actions on Facebook's database~\cite{stein_2011facebook}.

\paragraph{Report Users.} OSN operators can attempt to protect \s{young} children and teenage users from harassment by adding an option to report abuse or policy violations by other users in the network~\cite{facebook-report}. In some countries, social networks like Facebook and Bebo~\cite{bebo} have also added a ``Panic Button'' to better protect children~\cite{mashable}.

\subsection{Commercial Solutions}
\label{sec:industry}

Various commercial companies have expanded their traditional Internet security options and now offer software solutions specifically  for OSN users to better protect themselves against threats. In this section, we present mainstream software and application-protection solutions which were developed by well-known security companies\s{,} such as Symantec and Check Point, as well as solutions which were created by several startup companies\s{,} such as \s{Online Permissions 
Technologies, and open-source solutions, such as NoScript Security Suite. }

\paragraph{Internet Security Solutions.} Many security companies, such as AVG, Avira, Kaspersky, Panda, McAfee, and Symantec ~\cite{facebookav}, offer OSN users Internet security solutions. These software suites typically include anti-virus, firewall, and other Internet protection layers which assist OSN users in shielding their computers against threats such as malware, clickjacking, and phishing attacks.
For example, McAfee Internet Security software~\cite{macfeeinternet} provides its users with protection against various threats such as malware, botnet, and inappropriate sites. 

\s{
\paragraph{AVG PrivacyFix.} AVG PrivacyFix~\cite{privacyfix} is software available as a mobile application or a web browser add-on which offers its users a simple way to manage their privacy settings on Facebook, LinkedIn, and Google. Additionally, PrivacyFix helps its users block over 1,200 trackers by following their movements online.
The software also tells its users how much revenue they are generating for Facebook and Google.
}

\s{
\paragraph{FB Phishing Protector.} FB Phishing Protector~\cite{fbphishing} is a Firefox add-on which warns Facebook users when a suspicious activity is detected, such as a script-injection attempt. This add-on provides protection against various phishing attacks. 
}
\paragraph{Norton Safe Web.} Symantec's Norton Safe Web~\cite{safeweb} is a Facebook application with more than 500,000 users. It scans the Facebook user's News Feed and warns the user about unsafe links and sites.

\s{
\paragraph{McAfee Social Protection.} McAfee Social Protection~\cite{mcafeesp} is a mobile application which enables Facebook users to safeguard their uploaded photos by letting users control precisely who can view and download their images.
}

\s{
\paragraph{MyPermissions.} Online Permissions Technologies's MyPermissions~\cite{mypermmisions} is a web service that provides its users with convenient links to the permissions pages for many OSNs, such as Facebook, Twitter, and LinkedIn. 
These links can help users view and revoke the permissions they had given in the past to various applications, thus better protecting their privacy. Additionally, MyPermissions offers periodic email reminders that prompt users to check their OSN permissions settings. 
}

\s{
\paragraph{NoScript Security Suite.} NoScript Security Suite~\cite{noscript} is an open-source extension to  Mozilla-based web browsers like Firefox,  which allows executable web content such as JavaScript, Java, and Flash  to run only from trusted domains of the user's choice.  Blocking executable web  content  running from untrusted sites can protect OSN users from clickjacking and XSS attacks.
}

\paragraph{Privacy Scanner for Facebook.} Trend Micro's Privacy Scanner for Facebook~\cite{trendmicro} is an Android application which scans the user's privacy settings and identifies risky settings which may lead to privacy concerns. It then assists the user in fixing the settings.

\paragraph{Defensio.}  Websense's Defensio web service~\cite{defensio} helps protect social network users from threats like links to malware that \s{could} be posted on the user's Facebook page. The Defensio service also assists in \s{preventing information leakage by} controlling the user's published content \s{by} removing certain words from posts or filtering specific comments. 

\paragraph{ZoneAlarm Privacy Scan.} Check Point's  ZoneAlarm Privacy Scan~\cite{zonealarm_privacy} is a Facebook application which scans recent activity in the user's Facebook account to identify privacy concerns and to control what others can see. For instance, ZoneAlarm Privacy Scan can identify posts that expose the user's private information.

\paragraph{Net Nanny.}  ContentWatch's  Net Nanny~\cite{netnanny} is software which assists parents in  protecting their children from harmful content. Net Nanny lets parents monitor their children's social media activity on different \s{OSN} websites, such as Facebook, Twitter, and Flickr~\cite{flickr}. 

\paragraph{MinorMonitor.} Infoglide's MinorMonitor~\cite{minormonitor} is a parental control web service which gives parents a quick dashboard view of their child's Facebook activities and online friends. By using MinorMonitor, parents can be \s{informed} about questionable content that may have been revealed to their child, \s{and} they can identify over-age friends in their child's Facebook friends list.

\begin{table}

    \caption{\textbf{\s{Commercial Solutions Overview.}}}

	\begin{center}
  \begin{tabular}{c}

      \includegraphics[width=\textwidth]{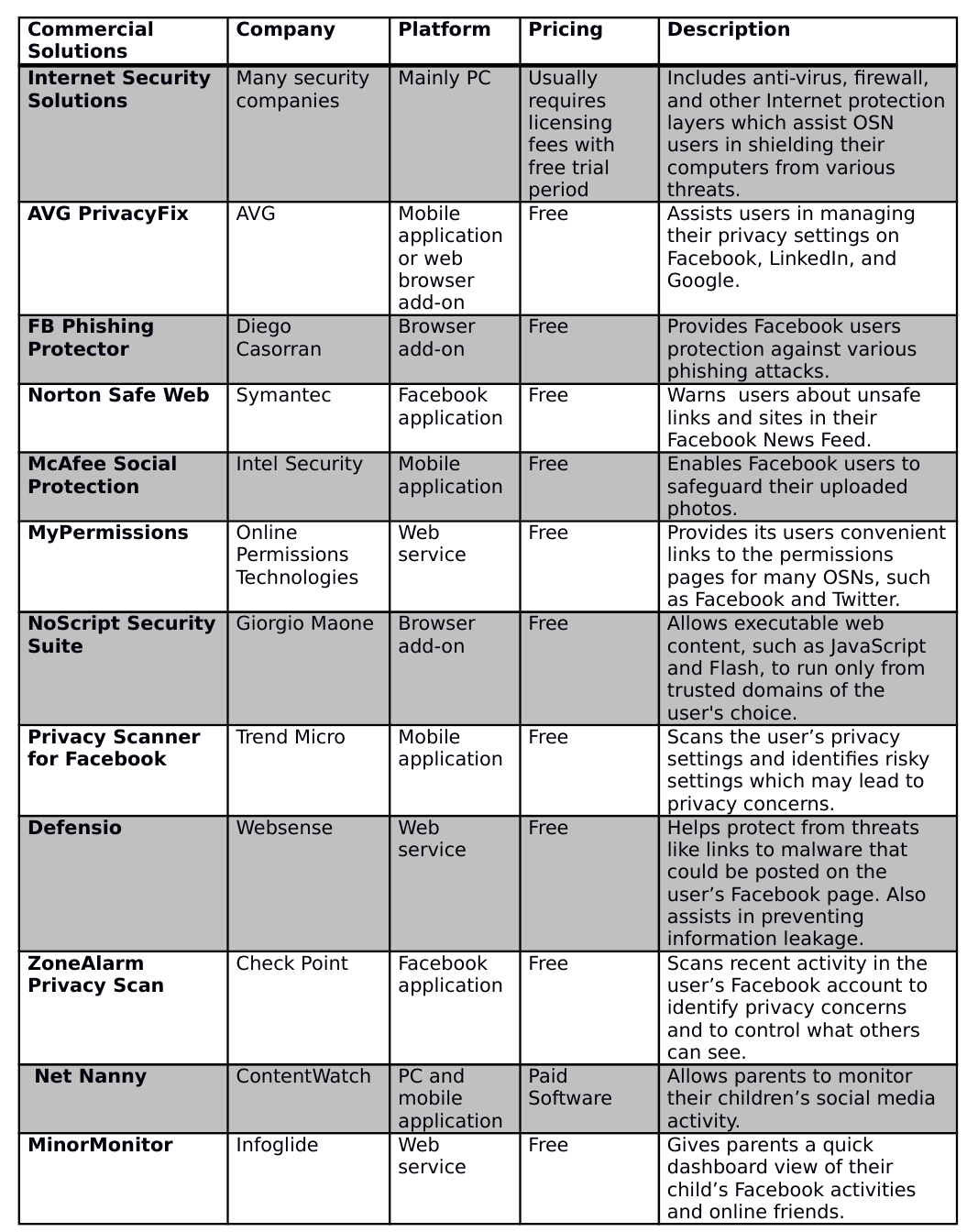}

  \end{tabular}
	\end{center}
	    \label{tab:commercial}

\end{table}

\subsection{Academic Solutions}
\label{sec:academic}

Several recently published studies have proposed solutions to various \s{OSN} threats. These solutions have primarily focused on identifying malicious users and applications. In this section, we present studies which provide solutions for \s{improving OSN users' privacy settings; for detecting phishing, spammers, cloned and fake profiles, and socware; and for preventing information and location leakage}.\footnote{Many of these solutions overlap and can assist in preventing more than one threat. For example, algorithms for identifying fake profiles can also help identify spammers and  phishing attacks.} \s{These academic solutions 
provide cutting-edge insight into dealing with social network threats. They
can be used by OSN operators to improve their users' security and privacy, by security companies to offer the customers better OSN protection, or by early-adopter OSN users who want to better protect themselves.}

\s{
\paragraph{Improving Privacy Setting Interfaces.} In recent years several studies have offered OSN users methods and applications to help them better understand and improve their social network privacy settings. 
In 2008, Lipford et al.~\cite{lipford2008understanding} introduced the Audience View interface for Facebook which enables users to view their profiles from the point of view of other Facebook users, whether from the point of view of a friend or that of a complete stranger. This type of interface can help OSN users know exactly which personal details are visible to other users and then change their privacy settings accordingly. In 2010, Fang and LeFevre~\cite{fang2010privacy} presented a template for the design of
a social networking privacy wizard for OSNs to automatically  configure the user’s privacy settings with minimal effort from the user. Fang and LeFevre also presented a sample privacy wizard based on their generic template. The sample wizard used active learning algorithms and was found to be ``quite effective in reducing the
amount of user effort, while still producing high-accuracy
settings''~\cite{fang2010privacy}. In 2012, Fire et al.~\cite{privacyprotector} presented The Social Privacy Protector add-on which can assist Facebook users in adjusting their privacy settings with just one simple click, according to predefined various privacy setting usage templates. Also in 2012, Paul et al.~\cite{Paul:2012:CCP:2187980.2188139} offered the C4PS privacy interface which utilizes simple principles of color coding
to highlight each attribute in the user's profile with a  particular color, depending on the group of people who have access to this attribute. 
Moreover, the interface enables users to change privacy settings for a specific attribute by simply clicking on buttons located near the specific attribute.

}
\paragraph{Phishing Detection.} Many researchers have suggested anti-phishing methods to identify and prevent phishing attacks; most of these methods have been based on techniques \s{that} attempt to identify phishing websites and phishing URLs~\cite{garera2007framework,ma2009beyond,xiang2011cantina}. \s{With} the increasing number of phishing attacks on OSNs~\cite{microsoft-report}, several researchers have suggested dedicated solutions for identifying \s{social network} phishing attacks. In 2012, Lee et al.~\cite{lee2012warningbird} introduced WarningBird, a suspicious URL detection system for Twitter
which can handle phishing attacks that conceal themselves by using conditional redirection URLs.  
Later in the same year, Aggarwal et al.~\cite{aggarwal2012phishari} presented the PhishAri technique, which can detect \s{whether or not} a tweet posted with a URL is phishing by utilizing specific Twitter features such as the account age and the  number of followers of the user who posted the suspicious tweet. 
 
\paragraph{Spammer Detection.} Many researchers have recently proposed solutions for spammer detection in \s{OSNs. }
In 2009, Benevenuto et al.~\cite{benevenuto2009detecting} offered algorithms for detecting video spammers which succeeded in 
identifying spammers among YouTube~\cite{youtube} users.
In 2010, \s{DeBarr} and Wechsler~\cite{debarr} used the graph centrality measure to predict \s{if} a user is likely to send spam messages. Wang~\cite{wang} proposed a method to classify spammers on Twitter by using content and social network graph properties. Stringhini et al.~\cite{stringhini} created more than 300  fake profiles (also referred \s{to} as ``honey-profiles'') \s{on Twitter}, Facebook, and MySpace  and successfully identified  spammers who sent spam messages to the fake profiles. Lee et al.~\cite{lee} also presented a method for detecting social spammers of different types by using honeypots combined with machine learning algorithms. In 2013, Aggarwal et al.~\cite{aggarwal2013detection} presented machine learning algorithms for detecting various type of spammers in Foursquare. Recently, Bhat and Abulaish~\cite{bhatcommunity} introduced a community-based framework to identify OSN spammers. \s{Also, Verma et al.~\cite{verma2014} presented a survey which reviews existing techniques for detecting spam users on Twitter.}

\paragraph{Cloned Profile Detection.} In 2011, Kontaxis et al.~\cite{kontaxis} proposed a methodology for detecting social network profile cloning. They designed and implemented a prototype which can be employed to investigate \s{whether or not} users have fallen victim to clone attacks. In 2013, Shan et al.~\cite{shan2013enhancing}  presented the CloneSpotter which can be deployed into the OSN infrastructure and \s{can} detect cloning attacks by using  users' data records, such as a user's login IP records that are available to the OSN operator.

\paragraph{Fake Profile Detection.} In recent years, researchers have \s{developed} algorithms, techniques, and tools  to identify fake profiles  and prevent various sybil attacks \s{via OSNs}.\footnote{Although the common goal of both fake profile algorithms and sybil defense algorithms is to identify fake profiles, a \s{difference exists: Fake} profile detection algorithms \s{seek} to identify fake profiles in general, including cases of cyber predators which hold only a few fake profiles in the OSN\s{;} sybil defense algorithms  are  a private case of fake profile detection algorithms and are usually intended to identify attackers \s{who} create  a large number of fake profiles in the OSN.}
In 2006, Yu et al.~\cite{yu2006sybilguard} presented the SybilGuard decentralized protocol that  assists in preventing sybil attacks. 
Later, in 2008, Yu el al.~\cite{yu2008sybillimit} also presented the SybilLimit protocol, a near-optimal defense against sybil attacks using social networks. 
In 2009, Danezis and Mittal~\cite{danezis2009sybilinfer} offered the SybilInfer defense algorithm which can distinguish between ``honest'' and ``dishonest'' users.  In the same year, Tran et al.~\cite{tran2009sybil} presented the SumUp sybil defense \s{system} to limit the number of fake votes cast by sybils. 

In 2012, Cao et al.~\cite{cao2012aiding} introduced the SybilRank tool which utilizes OSN graph properties to rank users according to their perceived likelihood of being fake. Later, they deployed SybilRank in the operation center of Tuenti~\cite{tuenti}, the largest OSN in Spain, and estimated that about 90\% of the 200,000 users \s{who} received the lowest \s{rank} were actually fake profiles. 
In the same year, Wang et al.~\cite{wang2012social} proposed a crowdsourced fake profiles detection system and evaluated it using data from Facebook and from Renren~\cite{renren}, a Chinese \s{OSN}.
\s{Also,} in 2012, Fire et al.~\cite{fire2012strangers} presented an algorithm for identifying malicious profiles using the social network's own topological features. They evaluated their methods on three directed \s{OSNs} \textemdash  Academia.edu~\cite{academia}, Anybeat,\footnote{\s{As of May 2012, the Anybeat OSN has been shutdown.}} and Google+ \textemdash  and succeeded in identifying fake profiles and spammers. Fire et al.~\cite{privacyprotector} also presented The Social Privacy Protector application which assists Facebook users in identifying fake profiles among their friends. They used the dataset created by The Social Privacy \s{Protector} application and developed machine learning classifiers which can identify fake profiles \s{on} Facebook~\cite{fire2013friend}.
Recently, Wang et al.~\cite{wang2013you} presented a system which can detect fake profiles based on analyzing clickstream models.
\s{Additional surveys regarding solutions to sybil attacks have also been presented by Levine et al.~\cite{levine2006survey} and by Hoffman et al.~\cite{hoffman2009survey}.}

\paragraph{Socware Detection.} In the last few years, \s{several} studies have tried to better understand and identify socware.
In 2012, Rahman et al.~\cite{rahman2012efficient} presented the MyPageKeeper Facebook application \s{that} aims to protect Facebook users from damaging posts on \s{their timelines}. Rahman et al. also presented Facebook’s Rigorous Application Evaluator (FRAppE) for detecting malicious applications on Facebook~\cite{rahman2012frappe}. In 2013, Huang et al.~\cite{huang2013analysis} studied the socware ecosystem and discovered several insights about socware propagation characteristics that can assist in future research on the detection and prevention of socware propagation.

\s{
\paragraph{Preventing Information and Location Leakage.} 
In their study on privacy leaks \s{on} Twitter, Mao et al.~\cite{mao2011loose} offered a ``guardian angel service''  that can monitor users' tweets and alert users to potential privacy violations. Their offered solution can be based on classifiers they constructed throughout their study which can identify tweets containing private information, such as vacation plans. Moreover, G{\'o}mez-Hidalgo et al.~\cite{gomez2010data} used Named Entity Recognition (NER) algorithms to prevent data leakage. In their study, they implemented a prototype to demonstrate how their methods can prevent data leakage.  Their methods may also be used to prevent OSN users from exposing their locations.
Recently, Ghiglieri et al.~\cite{personaldlp2014} presented the Personal DLP tool to help OSN users  better understand and evaluate the sensitivity of their posted statuses.
The study included 221 participants, and \s{the developed Personal DLP prototype was found to have a positive impact on users' privacy awareness}.
}

\begin{table}[hp]

    \caption{\textbf{\s{Online Social Network Threats and Their Corresponding Solutions }}}

	\begin{center}

  \begin{tabular}{c}

      \includegraphics[width=\textwidth]{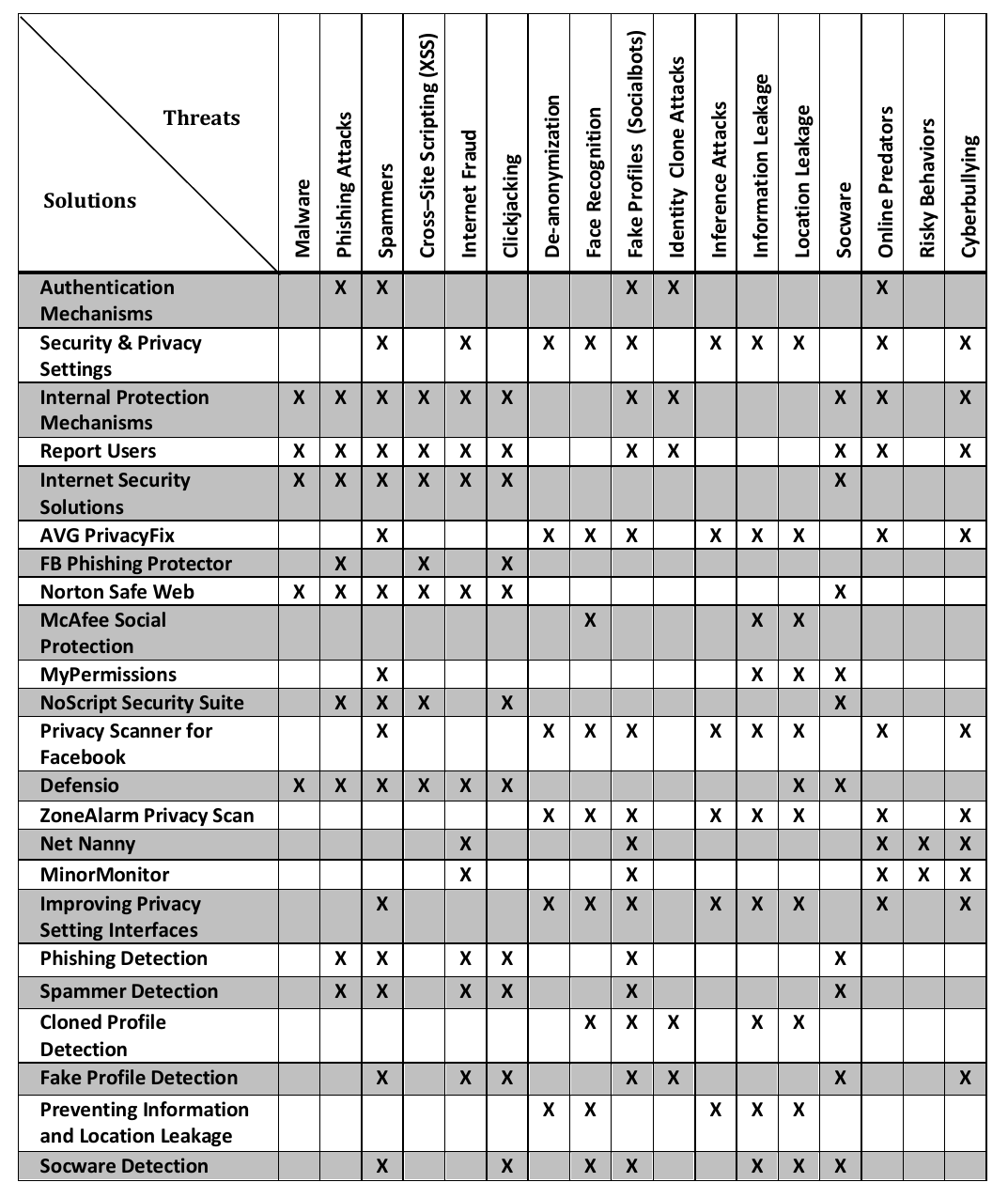}

  \end{tabular}
	\end{center}
	    \label{tab:threats_and_solutions}

\end{table}
\s{
\section{Discussion}
\label{sec:disc} 
In Section~\ref{sec:threats}, we presented the many threats that \s{can} jeopardize OSN users' security and privacy. These threats attempt to achieve one or more of the following goals: (a) gain access to the user's resources, such as passwords and credit card numbers (see Section~\ref{sec:classical}); (b) gain access to the user's private and sensitive information, such as age, political views, and current or future whereabouts (see Section~\ref{sec:modern}); (c) utilize the gained control over the user's OSN profile as a spreading platform to attack his or her trusting online friends; and (d) locate future potential victims (see Sections~\ref{sec:modern} and~\ref{sec:children}). 
Some of these threats are passive; they use only the user's lack of awareness or knowledge to achieve their goals. For example, the face recognition threat introduced in Section~\ref{sec:modern} can simply utilize the user's public profile photos to create a biometric database. 
Other threats are active, and their goal is to try and set up the users. 
For example, the clickjacking threat tries to trick OSN users into clicking on something different from what they had intended to click (see Section~\ref{sec:modern}).
Alarmingly, many of the presented threats are not limited to cyberspace but  have the potential to threaten the user's well-being in the real world as well. 
\s{For example, 
it has been suggested that most burglars use OSNs such as  Facebook and Twitter to target their victims~\cite{burglars2011}.}

To better protect OSN users from the above mentioned threats, OSN operators, commercial security companies,  and academic researchers offer OSN users a variety of security and privacy solutions which are presented in Section~\ref{sec:solutions}.  
Similar to real-world security solutions, these solutions can provide OSN users with several layers of protection against these threats. 
The first protection layer, which parallels the functionality of a \textit{door lock}, strives to prevent unwelcome intruders from entering and viewing OSN users'  personal posts and details. This layer consists of different security and privacy settings offered by various OSN operators. However, in many cases the average OSN user does not know or is unaware of the best way to ``lock'' his or her profile, instead  leaving the privacy settings on default, which often provides insufficient protection~\cite{fire2013friend,krishnamurthy2009leakage}. 
To assist such users, security companies and academic researchers have developed solutions, such as  \s{Privacy Scanner for Facebook~\cite{trendmicro}, ZoneAlarm Privacy Scan~\cite{zonealarm_privacy}}, and The Social Privacy Protector~\cite{privacyprotector}, all of which can assist OSN users in improving their privacy settings. However, much like in real life, sometimes OSN users can forget to ``lock their door,'' and consequently they may leak sensitive information about themselves, such as their future vacation plans or their medical condition~\cite{mao2011loose}. To prevent this type of exposure, researchers~\cite{Cheng:2010:YYT:1871437.1871535,mao2011loose} and security companies~\cite{defensio}  have offered solutions that automatically scan the users' posted information and prevent them from uploading posts that contain their sensitive information. 

The second protection layer parallels the functionality of a \textit{security alarm}, and it aims to prevent  malicious users from collecting OSN users' personal posts and details, that is, to prevent these malicious users from hacking into the innocent users' devices and social network accounts. This layer consists of the different
commercial Internet security solutions (see Section~\ref{sec:industry}), as well as the various phishing, fake profile, and socware detection solutions offered by academic researchers that the OSN users can install by themselves (see Section~\ref{sec:academic}). 
These types of solutions can be very effective in identifying  active threats, which in many cases attempt to infect as many OSN users as possible. In most cases, however, these solutions are insufficient for identifying more targeted threats, such as de-anonymization attacks, identity clone attacks, inference attacks,  and online predators, all of which choose to target individuals using an OSN.

The third protection layer, which functions as a \textit{security camera}, is a special layer specific to children and their OSN use. This layer aims to protect both young children and teenagers by enabling parents to monitor online activity primarily via various monitoring software such as Net Nanny~\cite{netnanny} and MinorMonitor~\cite{minormonitor}. This solution can help parents protect their children from targeted threats such as online predators and cyberbullying.

The fourth protection layer, which can be likened to the functionality of a \textit{neighborhood watch}, uses wisdom of the crowd to pinpoint malicious users in the OSN. This layer consists of various solutions such as the option to report other social network users to an OSN operator. OSN users can work together to identify  threats such as fake profiles, clickjacking, internet fraud, socware, and cyberbullying, and report them to the OSN operator.

The fifth protection layer, which parallels the functionality of a \textit{police force}, includes authentication mechanisms which are responsible for making sure that only real people can log into the OSN. The authentication mechanisms can assist in identifying malicious users, such as socialbots, and prevent them from logging into the OSN and attacking other social network users.  
Additionally, due to its almost unlimited access to OSN users' data, metadata, and activities, the OSN operator can identify many potential threats based on the full social network topology, along with users' IP addresses, login times, and behavioral patterns, which in most cases are accessible only to the OSN operator.
Moreover, as demonstrated in Sections~\ref{sec:operator} and~\ref{sec:academic}, utilizing these unique datasets can help protect OSN users from threats such as  phishing attacks~\cite{aggarwal2012phishari}, spammers~\cite{stringhini}, cloning attacks~\cite{shan2013enhancing},  and fake profiles~\cite{fire2012strangers}. 
\s{Fire et al.~\cite{fire2012strangers} showed how the OSN operator can utilize the full social network graph topology in order to identify fake profiles and spammers.  
Furthermore, as demonstrated by Stringhini et al.~\cite{stringhini} the OSN operator can use its control over the network to scatter 
many ``honey-profiles'' that can assist in identifying malicious users, such as spammers.}

These five protection layers can give OSN users sufficient protection against almost all of the threats described in Section~\ref{sec:threats} (also see Table~\ref{tab:threats_and_solutions}). 
Moreover,  if the OSN users choose to enable only the first three protection layers, they are still safeguarded from most of the described threats. 
Nevertheless, OSN operators \textemdash due to  their control of the network, their unique access to all  users' data and metadata, and their ability to monitor users' activities OSN operators\textemdash are in the best position to improve their users' security and privacy.

}

\section{Recommendations}
\label{sec:recom}
\begin{figure}[thb]

\begin{center}
\includegraphics[width=0.95\textwidth,clip]{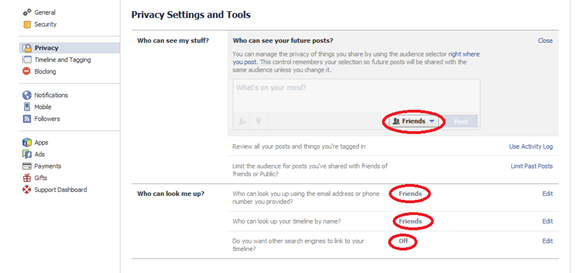}
\end{center}
\caption{\textbf{An Example of Recommended Privacy Settings on Facebook.} Only friends can have access to the user's private information.}
\label{fig:privacy}

\end{figure}

As we have demonstrated throughout this study, \s{OSN} users are facing prevalent and varied security and privacy threats. 
Fortunately, \s{there are} many software solutions and techniques that exist today which can assist OSN users \s{in better defending} themselves against these threats. 
In this section, we provide several easy-to-apply methods which can help OSN users improve their security and privacy in social networks such as Facebook and Twitter. We advise OSN users who want to better protect themselves in these platforms to implement the following eight recommendations in each of their OSN accounts:

\begin{enumerate}
\item \textbf{Remove Unnecessary Personal Information.} We advise OSN users to review the details they have inserted into their \s{OSN} accounts and remove extraneous information about themselves, their family, and their friends. It is also recommended that users hide their friends list if possible, to prevent inference attacks. 
\s{Additionally, we advise users not to use their full name when using OSNs, and in order to prevent face recognition, we highly recommend users not to use an identifiable image as their profile picture. }

\item \textbf{Adjust Privacy and Security Settings.} In many social networks, like Facebook, the default privacy settings are insufficient. \s{Yet} a recent study \s{has} showed that many Facebook users tend to stay with their default privacy settings~\cite{fire2013friend}. In order for users to better protect themselves on Facebook and in other \s{OSNs}, we recommend modifying the privacy settings so that users' personal data will be exposed only to themselves, or at most to their friends only (for example, see Fig.~\ref{fig:privacy}). Additionally, if possible, we advise users to activate the secure browsing option and any other available authentication mechanisms (see Section~\ref{sec:solutions}), such as Twitter's two-factor authentication~\cite{twitter2factors}.

\item \textbf{Do Not Accept Friend Requests From Strangers.} As we demonstrated in Section~\ref{sec:threats}, fake profiles are quite common and often dangerous. Therefore, if a user receives a friend request from an unknown person, we recommend ignoring such a request. If the user is uncertain and is considering approving the friend request, we recommend performing a short background check on the new ``friend'' and, at a minimum, insert the friend's profile image into Google Images \s{search}~\cite{googleimg} and submit the friend's full name and other details to other search engines in order to validate the authenticity of the individual.  
In order to identify and remove strangers who are already listed as friends with the user, we recommend OSN users examine their friends list or use applications such as The Social Privacy Protector~\cite{privacyprotector} and periodically remove friends with whom they are not familiar or \s{friends} who should not have access to personal information.

\item \textbf{Install Internet Security Software.} We advise OSN users to install at least one of the many commercial Internet security software products; Facebook offers several free security downloads~\cite{facebookav}. We also encourage users to install other security and privacy products as described in Section~\ref{sec:industry}.

\item \textbf{Remove Installed Third-Party Applications.}  Unbeknown to many users, third-party applications \s{frequently collect} online personal data. A recent study showed that 30\% of an examined group of Facebook users had at least forty applications installed on their accounts~\cite{fire2013friend}. It is recommended that \s{users do} not install new, unnecessary applications on \s{their accounts}. Moreover, users are advised to \s{periodically} go over their list of installed applications and remove any unnecessary applications.
\s{
\item \textbf{Do Not Publish Your Location.} As we described in Section~\ref{sec:modern}, many users publish their current or future location in multiple OSNs, and this information  can be used by criminals or stalkers. It is recommended that users avoid publishing any geographic location whatsoever in their accounts. Moreover, users are advised to disable geotagging on their mobile devices and cameras to prevent uploading of photos and videos that may contain location information.}

\item \textbf{Do Not Trust Your OSN Friends.} As we described in Section~\ref{sec:threats}, \s{OSN}
users tend to trust their friends in the social network.   \s{Since this trust can be misplaced}, we recommend OSN users take extra precautions when communicating with their online friends. We also recommend  \s{that users} think twice before offering any personal and sensitive information about themselves, even when posting photos. OSN users should definitely avoid revealing their home address, phone number, or credit cards numbers.

\s{
\item \textbf{Monitor Your Children's OSN Activity.}  We strongly advise parents to apply all the above mentioned recommendations to their children's OSN profiles. Additionally, we  recommend parents  monitor their children's online activity in OSNs. This monitoring can be done manually or by using one of the monitoring software products which we reviewed at the end of Section~\ref{sec:industry}. Moreover, we highly recommend that parents and their children periodically scan the friends list together in order to remove unwelcome ``friends.''  
}

\end{enumerate}

\section{Future Research Directions}
\label{sec:future}
The field of \s{OSN} security and privacy is a new and emerging one, \s{offering} many directions to \s{pursue}. Security researchers can continually provide better solutions to online threats; they can also discover new security threats to address. We believe that in order to improve the present solutions, the next step is to create synergy among the different security solutions which were presented in Section~\ref{sec:academic}. This will create more robust and effective security solutions for detecting fake profiles, spammers, phishing attacks, socware, and other threats.

Besides the creation of synergy, another \s{worthwhile direction is to apply various algorithms to enhance OSN security. A variety of} Natural Language Processing (NLP) techniques and temporal analysis algorithms \s{can be utilized;} combining \s{these} with existing solutions \s{would} provide better and more accurate protection against \s{social network threats}. For example, researchers can predict many users' private traits, such as age and gender, based on their Facebook likes~\cite{kosinski2013private}. Combining this algorithm with other topological-based fake profile detection methods (see Section~\ref{sec:academic}) can assist in spotting \s{phony} details, such as a false age, thus identifying fake profiles.  
\s{Other algorithms also can be utilized: Various Data Leak Prevention (DLP) algorithms can analyze and monitor OSN users' posted information, recommending to the users which of their posted information might be sensitive and therefore advised to be removed from social network.  Additionally, state-of-the-art anomaly detection algorithms could be used to develop solutions for identifying fake OSN user accounts or OSN user accounts that have been compromised. 

A further research direction for improving OSN users’ privacy is to analyze and evaluate the different existing privacy solutions offered by OSN operators, pinpointing their shortcomings and suggesting methods for improving privacy solutions. Research that develops techniques to better educate users about these solutions would also be of value, as would techniques to make users more aware of existing OSN threats.
}

\s{Additional possible future research directions include developing  privacy-preserving OSNs such as Safebook~\cite{cutillo2009safebook}, and developing solutions for privacy-preserving ad hoc social networks (i.e. self-configuring social networks that connect users using mobile devices~\cite{li2009mobisn}), such as the semantics-based mobile social network (SMSN) framework~\cite{li2012semantics}.
As SMSN grows in popularity, addressing security concerns will be increasingly important.
}

One additional possible future research direction includes studying the emerging security threats due to the increasing popularity of geo-location tagging of social network users~\cite{ruiz2011location} \s{in order to offer} solutions for threats \s{with geosocial specificity}. 

\section{Conclusions}
\label{sec:conclousions}
\s{OSNs} have become part of our everyday life and, on average, most Internet users spend more time on social networks than in any other online activity (see Section~\ref{osnusage}). We enjoy using \s{OSNs} to interact with other people through the sharing of experiences, pictures, \s{and videos.} Nevertheless, social networks have a dark side ripe with hackers, fraudsters, and online predators, all \s{of whom are} capable of using \s{OSNs} as a platform for procuring their \s{future victims}. In this paper, we have presented scenarios which threaten \s{OSN} users and can jeopardize their identities, privacy, and well-being \s{in both the virtual world as well as the real world} (see Section~\ref{sec:academic}). Furthermore, we have provided examples \s{of} many of the presented threats in order to demonstrate that these threats are real and can endanger every user. \s{We} have \s{also} emphasized certain threats which challenge the safety of \s{young} children and teenagers \s{across} the \s{OSN}  cyberspace.

There are remedies to these threats, and we have offered a range of
solutions which help protect an OSN user’s privacy and security (see
Section~\ref{sec:solutions}). \s{However, as demonstrated in Table~\ref{tab:threats_and_solutions}, the presented solutions are not magical antidotes that will provide full protection to a user’s privacy and security. In order
to be well protected against the various online threats, users must stay attentive to the information they post online, and they must employ more than one solution.
In many cases, the users should seek the OSN provider’s
assistance in providing tools (see Section~\ref{sec:operator}) both to better protect their
privacy and to identify potential threats.}

\s{
We have outlined eight recommendations that are simple to implement
for OSN users to better protect themselves (see Section~\ref{sec:recom}).
We advise OSN users to not only adopt our recommendations but also to educate themselves and their loved ones regarding online threats. All social network users must consider very carefully what personal information is being revealed about themselves, about their friends,
and about their workplaces. Users should also know that the information
they post in OSNs can be cross-referenced with other data sources~\cite{krishnamurthy2013privacy} and
could be used to infer their personal and intimate details. If a user’s personal information falls into the wrong hands, it could potentially cause a vast amount of damage, and
in many cases there is no way to recapture what has been lost.}

\s{In addition, parents must monitor their children’s activity in these social platforms. As parents, we cannot be naïve; we need to recognize the enticements of social networks and be aware of hidden dangers. We are obligated to educate our children to be aware of potential threats, and we must teach them not to engage with strangers either in the real world or in the cyber world. }


\s{
As far as future research (see Section~\ref{sec:future}), OSNs offer fertile ground for new and interesting research with many opportunities to pursue, such as improving the current state-of-the-art security products, discovering new types of security and privacy threats, and developing and evaluating new privacy solutions and schemes. Overall, researchers can play a significant role by recognizing the value of solution synergies and by applying useful techniques and algorithms. Social networks can enhance our lives, but we must take the correct precautions to preserve our security and privacy.}


\section*{Acknowledgment}
We would like to thank Jennifer Brill and Liza  Futerman for proofreading this article.
\s{Especially, we  want to thank Carol Teegarden for her editing expertise and endless helpful advice which guided this article to completion. }
\s{We  also want to thank 
the anonymous reviewers for their helpful comments.}





\bibliographystyle{abbrv}
\bibliography{threats_and_solutions}

\end{document}